\documentclass[cits,a4paper]{PoS}

\usepackage{amsmath,amssymb}
\usepackage{subfigure}
\usepackage{graphicx}

\newcommand{\mps}{m_\mathrm{PS}}
\newcommand{\fps}{f_\mathrm{PS}}

\newcommand{\preprint}{\newline%
  \begin{picture}(0,0)
  \put(115,120){\rm\small ROM2F/2008/23, DESY 08-139, SFB/CPP-08-78, LTH806, HU-EP-08/41}
  \end{picture}}

\graphicspath{{plots/}}

\title{Scaling and chiral extrapolation of pion mass and decay constant with
  maximally twisted mass QCD\preprint}

\ShortTitle{Pion mass and decay constant with maximally twisted mass QCD}

\author{Petros Dimopoulos, Roberto Frezzotti\\
  Dip. di Fisica, Universit\`a di Roma ``Tor Vergata''\\
  Via della Ricerca Scientifica 1, 00133 Rome, Italy\\
  E-mail: \email{\{dimopoulos,frezzotti\}@roma2.infn.it}}

\author{Gregorio Herdoiza, Karl Jansen\\
  DESY\\
  Platanenallee 6, 15738 Zeuthen, Germany\\
  E-mail: \email{\{Gregorio.Herdoiza,Karl.Jansen\}@desy.de}}

\author{Chris Michael\\
  Theoretical Physics Division, Department of Mathematical Sciences,
  University of Liverpool\\
  Liverpool L69 3BX, UK\\
  E-mail: \email{c.michael@liverpool.ac.uk}}

\author{\speaker{Carsten Urbach}\\
  Humboldt-Universit{\"a}t zu Berlin, Institut f{\"u}r Physik,\\
  Newtonstr. 15, 12489 Berlin, Germany\\
  E-mail: \email{Carsten.Urbach@physik.hu-berlin.de}}

\abstract{We present an update of the  results for pion mass and pion
  decay constant as obtained by the ETM collaboration in large scale
  simulations with maximally twisted mass fermions and two mass
  degenerate flavours of light quarks. We discuss the continuum,
  chiral and infinite volume extrapolation of these quantities as well
  as the extraction of low energy constants, and investigate possible
  systematic uncertainties.}

\FullConference{The XXVI International Symposium on Lattice Field Theory\\
		 July 14-19 2008\\
		 Williamsburg, Virginia, USA}

\begin{document}

\section{Introduction}

We present an update of the $n_f=2$ results obtained by the European
Twisted Mass collaboration (ETMC) for the pseudo scalar mass and decay
constant. A good understanding of continuum, thermodynamic and chiral
limits is essential in order to obtain reliable results, which
can eventually be compared to experiment. The physics of the light
pseudo scalar meson is a prime example for investigating these
extrapolations, because its mass and decay constant can be obtained
with high precision in lattice simulations and chiral perturbation
theory ($\chi$PT) is best understood for those two quantities. In
addition, such an investigation allows to extract other quantities of
phenomenological interest, such as low energy constants and quark masses.

First results for the pseudo scalar mass and decay constant obtained
for the large scale simulations of the ETM collaboration can be found
in Refs.~\cite{Boucaud:2007uk, Urbach:2007rt, Dimopoulos:2007qy,
  Boucaud:2008xu}.

\begin{table}[t!]
  \centering
  \begin{tabular*}{1.0\textwidth}{@{\extracolsep{\fill}}lccccccc}
    \hline\hline
    Ensemble & $L^3\times T$ & $\beta$ & $a\mu_q$ & $\kappa$ &
    $\tau_\mathrm{int}(P)$ & $\tau_\mathrm{int}(am_\mathrm{PS})$ & $\tau$\\
    \hline\hline
    $B_1$ & $24^3\times 48$ & $3.9$ & $0.0040$ & $0.160856$ & $47(15)$
    & $7(1)$ & $0.5$\\
    $B_2$ &  & & $0.0064$ &  & $23(7)$
    & $17(4)$ & $0.5$ \\
    $B_3$ &  & & $0.0085$ &  & $13(3)$ & $10(2)$ & $0.5$\\
    $B_4$ &  & & $0.0100$ &  & $15(4)$ & $7(2)$ & $0.5$\\
    $B_5$ &  & & $0.0150$ &  & $30(8)$ & $20(6)$ & $0.5$\\
    $B_6$ & $32^3\times 64$ & $3.9$ & $0.0040$ & $0.160856$ & $37(11)$
    & $2.8(3)$ & $0.5$ \\
    $B_7$ &  &  & $0.0030$ &  & $51(19)$ & $7(1)$ & $1.0$ \\
    \hline\hline
  \end{tabular*}
  \caption{Update of the ensembles produced with $\beta=3.9$ by the
    ETM collaboration. For the other $\beta$-values see table 1 of
    Ref.~\cite{Urbach:2007rt}. We give the lattice volume $L^3\times
    T$, the twisted mass parameter $a\mu_q$, the hopping parameter
    $\kappa=1/(8+2am_0)$ and the trajectory length $\tau$. In addition
    we provide values for the integrated autocorrelation time of two
    typical quantities, the plaquette $P$ and the pseudo scalar mass
    $am_\mathrm{PS}$, in units of $\tau=0.5$.}
  \label{tab:setup}
\end{table}

ETMC has generated large sets of gauge configurations
for different values of the coupling constant ($\beta=3.8$, $a\sim0.1\
\mathrm{fm}$; $\beta=3.9$, $a\sim0.085\ \mathrm{fm}$;
$\beta=4.05$, $a\sim0.065\ \mathrm{fm}$), for various volumes
($2.1-2.8\ \mathrm{fm}$) and a 
number of bare quark masses corresponding to pseudo scalar meson masses
ranging from $\sim 260$ to $\sim 700\ \mathrm{MeV}$. The list of
ensembles at $\beta=3.9$ can be found in table~\ref{tab:setup}, which
contains the newly generated ensemble $B_7$ corresponding to a pseudo
scalar meson mass of about $\mps\sim 265\ \mathrm{MeV}$. For the other
$\beta$-values we refer to table~1 of Ref.~\cite{Urbach:2007rt}. 

In the gauge sector we employ the so-called tree-level Symanzik improved
gauge action (tlSym) \cite{Weisz:1982zw}. The fermionic action for two
flavours of maximally twisted, mass degenerate quarks in the so called
twisted basis~\cite{Frezzotti:2000nk,Frezzotti:2003ni} reads
\begin{equation}
  \label{eq:sf}
  S_\mathrm{tm}\ =\ a^4\sum_x\left\{ \bar\chi(x)\left[ D[U] + m_0 +
      i\mu_q\gamma_5\tau^3\right]\chi(x)\right\}\, ,
\end{equation}
where $m_0$ is the untwisted bare quark mass tuned to its critical
value $m_\mathrm{crit}$, $\mu_q$ is the bare twisted quark mass,
$\tau^3$ is the third Pauli matrix acting in flavour space and $D[U]$ 
is the Wilson-Dirac operator.

At maximal twist, i.e.~$m_0=m_\mathrm{crit}$, physical observables are
automatically $\mathcal{O}(a)$ improved without the need to determine
any action or operator specific improvement
coefficients~\cite{Frezzotti:2003ni} (for a review see
Ref.~\cite{Shindler:2007vp}). With this being the main advantage, one
drawback of maximally twisted  mass fermions is that flavour symmetry
is broken explicitly at finite value of the lattice spacing, which
amounts to $\mathcal{O}(a^2)$ effects in physical observables, as will
be discussed later. Note that in the following we shall refer to the 
charged pseudo scalar meson mass as $\mps$ or $\mps^\pm$ and to the
neutral one as $\mps^0$.

For details on the set-up, tuning to maximal twist and the analysis
methods of the ETM collaboration we refer to
Refs.~\cite{Boucaud:2007uk, Urbach:2007rt, Boucaud:2008xu}. 
Recent results for light quark masses and decay constants, the light
baryon spectrum and the $\eta'$ meson are available in
Refs.~\cite{Blossier:2007vv,Alexandrou:2008tn} and
Ref.~\cite{Jansen:2008wv}, respectively. 
We shall \emph{only} consider
the ensembles at $\beta=3.90$ ($B$-ensembles) and $\beta=4.05$
($C$-ensembles, see table 1 of Ref.~\cite{Urbach:2007rt}) in this
proceeding contribution, because tuning to maximal twist at
$\beta=3.8$ was not sufficiently accurate at the lowest quark mass
values for the observables considered here.

\section{Results}

\subsection*{Flavour Breaking Effects}

\begin{figure}[t]
  \centering
  \subfigure[\label{fig:splitting}]%
  {\includegraphics[width=0.44\linewidth]{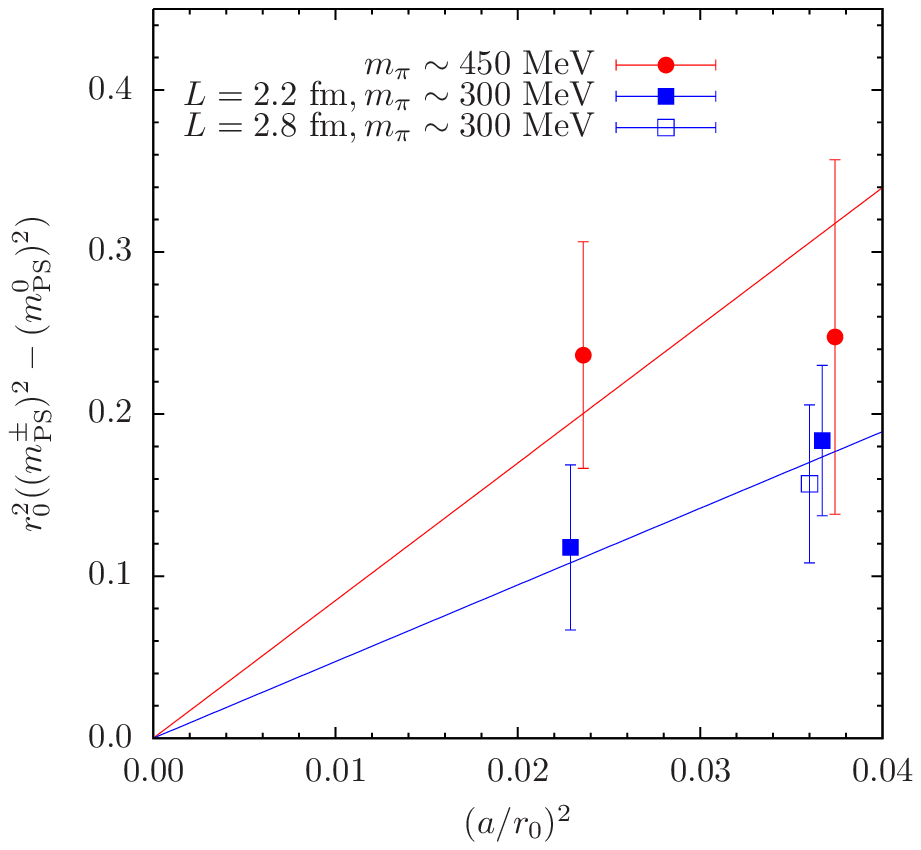}}
  \quad
  \subfigure[\label{fig:fps}]%
  {\includegraphics[width=0.46\linewidth]{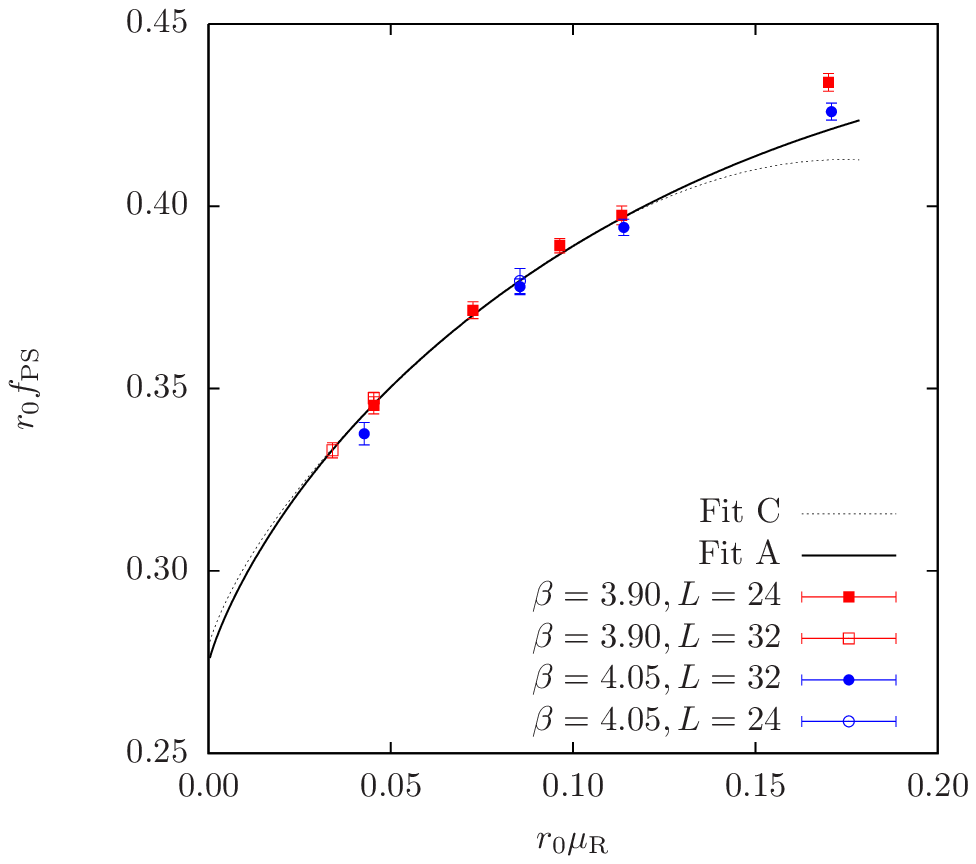}}
  \caption{(a) Mass splitting $r_0^2((\mps^\pm)^2-(\mps^0)^2)$ as a
    function of $(a/r_0)^2$. (b) Data for $r_0\fps$ as a  function of 
    $r_0\mu_R$ for $\beta=3.90$ and $\beta=4.05$ and Fits A and C.}
  \label{fig:latart}
\end{figure}

Flavour breaking effects have been investigated by ETMC for several
quantities. In figure~\ref{fig:splitting} we plot
$r_0^2((\mps^\pm)^2-(\mps^0)^2)$ as a function of $(a/r_0)^2$. It is
visible that mass splitting of the charged to neutral pseudo scalar
meson is large. However, the measured splittings are compatible
with being an $\mathcal{O}(a^2)$ effect, as expected, and they vanish
towards the continuum limit. 

All other possible splittings investigated so far are compatible with
zero. In table~\ref{tab:splitting} we have compiled the relative
difference $R_O=(O-O')/O$ for some selected simulation points and
observables $O$. Here $O$ ($O'$) denotes the charged (neutral)
quantity for mesons and $\Delta^+$ ($\Delta^{++}$) for baryons. The
values of $R_O$ are well compatible with zero for all 
observables $O$ besides the pion mass. However, some quantities, like
the vector meson decay constant $f_V$, are rather noisy, making
definite conclusions difficult. These results are compatible with a
theoretical investigation using the Symanzik effective
Langrangian~\cite{Frezzotti:2007qv}. 

\begin{table}[t]
  \centering
  \begin{tabular*}{0.8\linewidth}{@{\extracolsep{\fill}}lrrr}
    \hline\hline
     & $\beta$ & $a\mu_q$ & $R_O$ \\
     \hline\hline
     $af_\mathrm{PS}$ & $3.90$ & $0.004$ & $0.04(06)$ \\
    & $4.05$ & $0.003$ & $-0.03(06)$ \\ 
    $am_\mathrm{V}$  & $3.90$ & $0.004$ & $0.02(07)$ \\
    & $4.05$ & $0.003$ & $-0.10(11)$ \\
    $af_\mathrm{V}$  & $3.90$ & $0.004$ & $-0.07(18)$ \\
    & $4.05$ & $0.003$ & $-0.31(29)$ \\
    $am_\Delta$      & $3.90$ & $0.004$ & $0.022(29)$  \\
    & $4.05$ & $0.003$ & $-0.004(45)$ \\
    \hline\hline
  \end{tabular*}
  \caption{Comparison of some selected quantities for which an 
    isospin splitting can occur for twisted mass fermions. $R_O$ denotes the 
    measured relative size of the splitting.}
  \label{tab:splitting}
\end{table}

\subsection*{$\mathrm{SU}(2)$ $\chi$PT Fits}

As the details of finite size corrections for $\mps$ and $\fps$ were
discussed in Ref.~\cite{Urbach:2007rt} and the issue of continuum
extrapolation in fixed and finite volume for selected quantities in 
Refs.~\cite{Urbach:2007rt, Dimopoulos:2007qy}, we shall summarise here
only the main results: 
\begin{enumerate}
\item finite size effects in $\fps$ and $\mps$ can be described using
  chiral perturbation theory in the form of the resummed L{\"u}scher
  formula as described in Ref.~\cite{Colangelo:2006mp}. We denote the
  corresponding correction factors with $K_{f,m}^{\mathrm{CDH}}$,
  which depend among others on the low energy constants
  $\Lambda_{1-4}$.

\item within our current statistical precision lattice artifacts
  appear to be negligible, in particular for $\mps, \fps$, the
  quantities we consider here. In fixed volume and at
  fixed value of $r_0\mps$ the results 
  for $r_0\fps$ at $\beta=3.9$ and $\beta=4.05$ are always compatible
  with each other within our small statistical errors, and hence a constant
  extrapolation to the continuum limit seems justified. When we include
  also a linear term in $(a/r_0)^2$ in the extrapolation, a least
  square fit
  determines a value for the slope that is zero within
  errors. However, we include this effect into our systematic
  uncertainties. 
\end{enumerate}
We shall now present the results for a combined chiral, thermodynamic
and continuum extrapolation of $\mps$ and $\fps$ for the two
$\beta$-values $\beta=3.9$ and $\beta=4.05$. What we present here will
extend the results given in Refs.~\cite{Urbach:2007rt, Dimopoulos:2007qy}
by incorporating chirally extrapolated data for the renormalisation
constant $Z_\mathrm{P}$ and the Sommer parameter $r_0/a$ into the
fit. Details on the computation of $Z_\mathrm{P}$ (using the RIMOM)
and $r_0/a$ can be found in Refs.~\cite{Dimopoulos:2007fn, Boucaud:2008xu}. 

We perform combined fits to our data for $\fps$, $\mps$, $r_0/a$ and
$Z_P$ at the two values of $\beta$ with the formulae:
\begin{equation}
  \label{eq:fmps}
  \begin{split}
    r_0\fps &= r_0
    f_0\Bigl[1-2\xi\log\left(\frac{\chi_\mu}{\Lambda_4^2}\right) +
    D_{f_\mathrm{PS}}(a/r_0)^2 + T_f^\mathrm{NNLO}\Bigr]\
    K_f^\mathrm{CDH}(L)\, , \\
    (r_0 \mps)^2 &= \chi_\mu r_0^2\Bigl[
    1+\xi\log\left(\frac{\chi_\mu}{\Lambda_3^2}\right)+
    D_{m_\mathrm{PS}}(a/r_0)^2 + T_m^\mathrm{NNLO}\Bigr]\
    K_m^\mathrm{CDH}(L)^2\, ,\\
  \end{split}
\end{equation}
with $ \xi \equiv 2B_0\mu_R/(4\pi f_0)^2\ ,\chi_\mu\equiv
2B_0\mu_R\ , \mu_R \equiv \mu_q/Z_\mathrm{P}, f_0\equiv\sqrt{2} F_0$.
$T_{m,f}^\mathrm{NNLO}$ denote the continuum NNLO
terms~\cite{Leutwyler:2000hx}, which depend on $\Lambda_{1-4}$ and
$k_M$ and $k_F$, and $K_{m,f}^\mathrm{CDH}(L)$ the finite
size corrections~\cite{Colangelo:2006mp}. Based on the form of the
Symanzik expansion in the small quark mass region, we parametrise in
eq.~(\ref{eq:fmps}) the leading cut-off effects by the two
coefficients $D_{f_{\rm PS},m_{\rm PS}}$. 

\begin{figure}[t]
  \centering
  \subfigure[\label{fig:mps}]%
  {\includegraphics[width=0.45\linewidth]{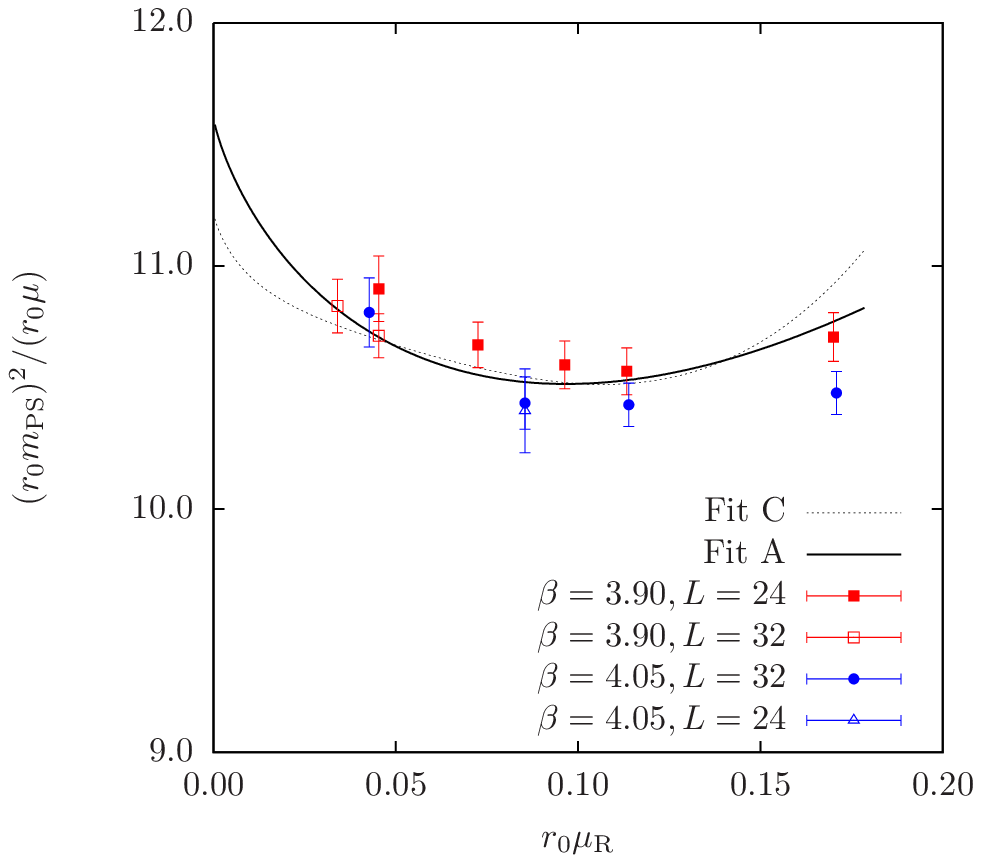}}
  \quad
  \subfigure[\label{fig:fpsasq}]%
  {\includegraphics[width=0.45\linewidth]{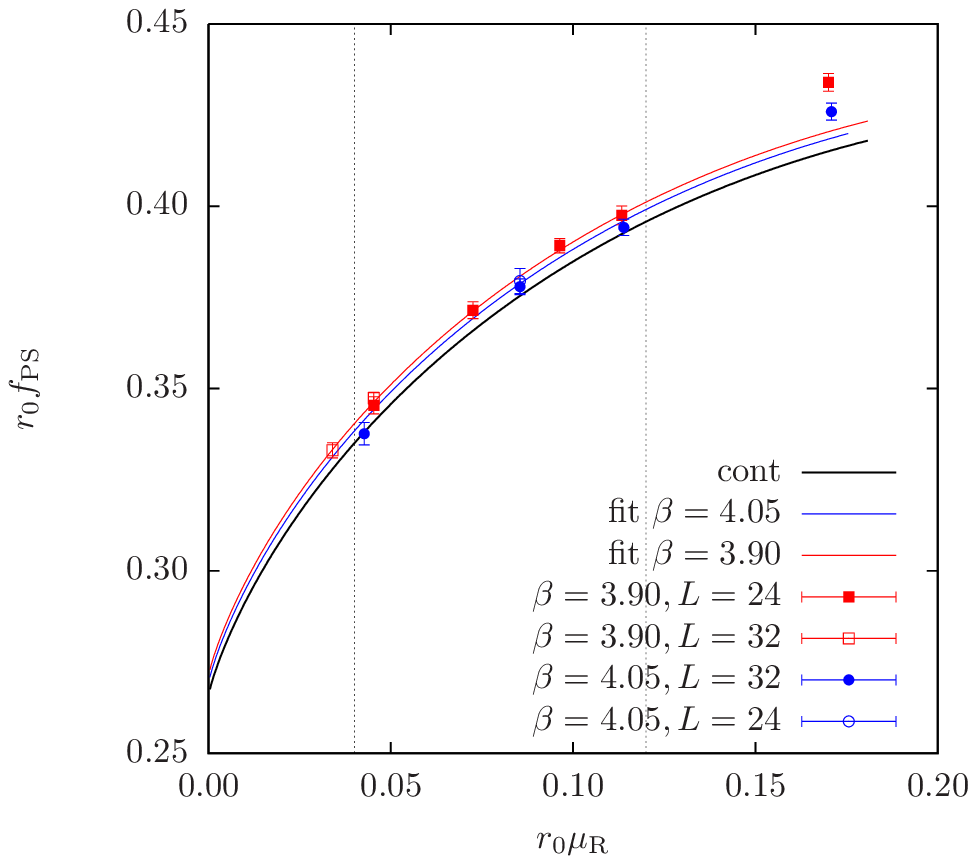}}
  \caption{(a) Data for $(r_0\mps)^2/r_0\mu_R$ as a function of
    $r_0\mu_r$ and Fits A and C. (b) Data for $r_0\fps$ as a
    function of $r_0\mu_R$ and resulting curves of Fit B. The vertical
    lines indicate the fit range.}
  \label{fig:comp}
\end{figure}

At NLO, i.e. setting $T_{m,f}^\mathrm{NNLO}\equiv0$, and neglecting finite size
corrections for the moment, there are the
following free parameters to be fitted to the data for $a\fps$,
$a\mps$, $r_0/a$ and $Z_\mathrm{P}$:
\[
r_0f_0,\ r_0B_0,\ r_0\Lambda_3,\ r_0\Lambda_4,\
\{r_0/a\}_\beta,\ \{Z_\mathrm{P}\}_\beta,\,\ D_{\mps},\ D_{\fps},
\]
where we indicate with the notation $\{...\}_\beta$ that there is one
parameter for each value $a$.

Finite size effects are corrected for by using the asymptotic formulae
from CDH, which is consistently included in the fit. However,
$K_{f,m}^\mathrm{CDH}$ depend on more parameters ($\Lambda_{1,2},
\tilde r_{1-4}$). Those we do not fit, but set them to the values
suggested in Ref.~\cite{Colangelo:2005gd} using the physical value of
$r_0$ as determined from the fit. This appears to be justified, since
we are able to describe our measured finite size effects.

When including NNLO terms into the fit there are four
additional parameters to be determined. We cannot fit them to the
data, because the fits become unstable. In order to be still able to
estimate systematic uncertainties from NNLO contributions, we include
priors for $r_0\Lambda_{1,2}$, $k_M$, $k_F$ into the fit when NNLO
terms are included. As priors we use for $k_{M,F}=0\pm1$ and for
$r_0\Lambda_{1,2}$ the values given in Ref.~\cite{Colangelo:2005gd}.

Our fit procedure can be viewed as first extrapolating the data to the
continuum limit and use continuum chiral perturbation theory
afterwards for the chiral and infinite volume extrapolations. 
For this reason we do not expect any influence of the neutral pseudo
scalar meson on the finite size effects: in the continuum all
three pseudo scalar mesons are degenerate. Note that setting
$D_{\mps,\fps}\equiv0$ corresponds to a constant continuum
extrapolation. Using the boostrap method to estimate the statistical
uncertainties, we performed the following fits 
\begin{enumerate}
\item Fit A: NLO continuum $\chi$PT, $T_{m,f}^\mathrm{NNLO}\equiv0$,
  $D_{\mps,\fps}\equiv0$, ensembles $B_{1,2,3,4,6}$ and $C_{1,2,3,5}$
  \vspace{-0.2cm}
\item Fit B: NLO continuum $\chi$PT, $T_{m,f}^\mathrm{NNLO}\equiv0$, $D_{\mps,\fps}$
  fitted, ensembles $B_{1,2,3,4,6}$ and $C_{1,2,3,5}$
  \vspace{-0.2cm}
\item Fit C: NNLO continuum $\chi$PT, $D_{\mps,\fps}\equiv0$,
  ensembles $B_{1,2,3,4,6}$ and $C_{1,2,3,5}$
  \vspace{-0.2cm}
\item Fit D: like Fit A, but ensembles $B_5$ and $C_4$ added
  \vspace{-0.2cm}
\item Fit E: like Fit A, but ensemble $B_7$ added
\end{enumerate}
Using the fitted parameters we can then determine low energy constants
like $\bar\ell_{3,4}$, the chiral condensate $\Sigma$ and the pseudo
scalar decay constant in the chiral limit $f_0$.

\subsection*{Discussion}

The fit results are summarised in table~\ref{tab:results}. In terms of
$\chi^2/\mathrm{dof}$ the Fits A, B, C and E provide a good
description of the data with $\chi^2/\mathrm{dof}\sim1$, whereas Fit
D, which includes simulation points with $\mps\sim600\ \mathrm{MeV}$,
has significantly larger $\chi^2/\mathrm{dof}$. We conclude from this
that $\chi$PT is not applicable for values of $\mps > 500\ \mathrm{MeV}$. 

To the contrary, including ensemble $B_7$ as in Fit E, and hence
extending the fit-range to a value of $\mps\sim265\ \mathrm{MeV}$
reveals completely consistent results with Fit A. This result makes us
confident that the extrapolation to the physical point is trustworthy.

Including lattice artifacts in the fit (Fit B) does change rather
little as compared to Fit A, and the coefficients $D_{\mps,\fps}$ are
compatible with zero, while the value of $\chi^2/\mathrm{dof}$ is not
significantly reduced: the differences between the results at
$\beta=3.9$ and $\beta=4.05$ can be explained with the variance
observed in $r_0/a$ and $Z_\mathrm{P}$. Hence, we have to reduce our
(already small) statistical errors even further to resolve
lattice artifacts in $\fps$ and $\mps$, indicating small lattice
artifacts in those two quantities.
When NNLO terms are included in the Fit (Fit C), the most significant
difference compared to Fit A is observed for $\bar\ell_3$. Though this
effect is not significant, we include it as a systematic error in our
final results.

These findings are visualised in figures~\ref{fig:fps}, \ref{fig:mps}
and \ref{fig:fpsasq}. In figure~\ref{fig:fps} we plot $r_0\fps$ as a
function of the renormalised quark mass $r_0\mu_R$ comparing Fits A
and C. In the range $0.04 \leq r_0\mu_R\leq 0.12$ the two fits
agree remarkably well, while for $r_0\mu_R>0.12$ both fail to
describe the data. Note that we might be seeing lattice artifacts of the order
$a^2\mu_q^2$ at these large masses, which would explain the
difference between the results at $\beta=3.9$ and $\beta=4.05$.
Similar conclusions can be drawn from figure~\ref{fig:mps}, where we
plot $(r_0\mps)^2/(r_0\mu_R)$ as a function of the renormalised quark
mass. 

In figure~\ref{fig:fpsasq} we show the result of Fit B for $r_0\fps$.
The three curves correspond to the fitted curve at $\beta=3.9$ (red),
the fitted curve at $\beta=4.05$ (blue) and to the continuum curve
(black). The differences between the three curves are rather small,
reflecting the result that $D_{\mps,\fps}$ are zero within errors.

\begin{table}[t]
  \centering
  \begin{tabular*}{1.\linewidth}{@{\extracolsep{\fill}}lrrrrr}
    \hline\hline
    $\Bigl.\Bigr.$ & Fit A & Fit B & Fit C & Fit D & Fit E\\
    \hline\hline
    $\bar\ell_3$ & $3.42(8)$ & $3.52(8)$ & $3.15(19)$ & $3.55(5)$ & $3.41(7)$
    \\
    $\bar\ell_4$ & $4.59(4)$ & $4.61(4)$ & $4.72(12)$ & $4.72(2)$ & $4.60(3)$ \\
    $\Sigma^{\overline{\mathrm{MS}}}(2\mathrm{GeV}) [\mathrm{MeV}^3]$ & $(-267(2))^3$ & $(-276(5))^3$ & $(-263(2))^3$
    & $(-269(1))^3$ & $(-267(1))^3$ \\
    $f_0\ [\mathrm{MeV}]$ & $121.66(7)$ & $121.6(1)$ & $121.7(3)$ & $121.39(5)$ &
    $121.64(7)$ \\
    $f_\pi/f_0$ & $1.0743(7)$ & $1.0746(9)$ & $1.0739(23)$ & $1.0767(4)$ & $1.0745(6)$ \\
    $D_{\mps}$ & -- & $-1.4(1.3)$ & -- & -- & --\\
    $D_{\fps}$ & -- & $+0.58(69)$ & -- & -- & --\\
    $\chi^2/\mathrm{dof}$ & $17.7/14$ & $12.9/12$ & $15.3/14$ & $46.7/18$ & $18.6/16$ \\
    \hline\hline
   \end{tabular*}
  \caption{Summary of fit results.}
  \label{tab:results}
\end{table}

\section{Conclusion and Outlook}

We have presented an update of the ETMC results for $\fps$ and $\mps$
and their continuum, thermodynamic and chiral extrapolations. The main
difference to the previous analysis in
Ref.~\cite{Urbach:2007rt,Dimopoulos:2007qy} is a new simulation point
at $\beta=3.9$ and the inclusion of $r_0/a$ and 
$Z_\mathrm{P}$ data into the fit. The main results are summarised with
$\bar\ell_3 = 3.42(8)(10)(27)$, $\bar\ell_4=4.59(4)(2)(13)$,
$\Sigma^{\overline{\mathrm{MS}}}(2\mathrm{GeV}) = (-267(2)(9)(4)\
\mathrm{MeV})^3$ and $f_\pi/f_0 = 1.0743(7)(3)(4)$. The first error is
statistical, the second estimates residual lattice artifacts and the
third effects from NNLO $\chi$PT. In addition we have presented
results indicating that flavour breaking effects are zero within the
statistical accuracy, with the exception of the neutral pseudo scalar
meson mass.

We thank all members of ETMC for the most enjoyable collaboration. 
This work has been supported in part by  the DFG
Sonder\-for\-schungs\-be\-reich/ Transregio SFB/TR9-03.

\bibliographystyle{h-physrev5}
\bibliography{bibliography}

\end{document}